\documentstyle{article}
         \def\ba{\begin{array}}
         \def\ea{\end{array}}
         \def\be{\begin{equation}}
         \def\bea{\begin{eqnarray}}
         \def\eea{\end{eqnarray}}
         \def\ee{\end{equation}}
         
         \def\d{{\rm d}}

\begin{document}
\begin{titlepage}
\vspace{-10mm}
\vspace{12pt}

\vskip 5 mm
\leftline{ \Large \bf
          Logarithmic $N=1$ superconformal field theories}
\vskip 1.5 cm
\leftline{\bf M. Khorrami$^{1,3,*}$, A. Aghamohammadi$^{2,3}$,
              A. M. Ghezelbash$^{2,3}$}
\vskip 5 mm
{\it
  \leftline{ $^1$ Department of Physics, Sharif University of Technology
             P.O. Box 11365-9161, Tehran, Iran. }
  \leftline{ $^2$ Department of Physics, Alzahra University,
             Tehran 19834, Iran. }
  \leftline{ $^3$ Institute for Studies in Theoretical Physics and
            Mathematics, P.O.Box  5531, Tehran 19395, Iran. }
  \leftline{ $^*$ mamwad@netware2.ipm.ac.ir}
  }
\vskip 4 cm
\begin{abstract}
We study the logarithmic superconformal field theories.
Explicitly, the two--point functions of $N=1$ logarithmic
superconformal field theories (LSCFT) when the Jordan blocks are two
(or more) dimensional, and when there are one (or more) Jordan block(s)
have been obtained. Using the well known three-point fuctions of $N=1$
superconformal field theory (SCFT), three--point functions of $N=1$ LSCFT
are obtained. The general form of $N=1$ SCFT's four--point functions is also
obtained, from which one can easily calculate four--point functions in $N=1$
LSCFT.

\end{abstract}
\vskip 10 mm
\end{titlepage}
\section {Introduction}
It has been shown by Gurarie \cite{Gu}, that conformal field theories
(CFT's) whose correlation functions exhibit logarithmic behaviour, can be
consistently defined. It is shown that if in the OPE of two local
fields, there exist at least two fields with the same conformal dimension, one
may find some special operators, known as logarithmic
operators. As discussed in \cite{Gu}, these operators with the ordinary
operators form the basis of a Jordan cell for the operators $L_i$.
In some interesting physical theories, for example dynamics
of polymers \cite{Sa}, the WZNW model on the $GL(1,1)$ super-group \cite{RS},
WZNW models at level 0 \cite{KM,CKLT,KL} , percolation \cite{Ca},
the Haldane-Rezayi quantum Hall state\cite{GFN}, and edge excitation in
fractional quantum Hall effect \cite{WWH}, one can  naturally find
logarithmic terms in correlators. Recently the role of logarithmic operators
have been considered in study of some physical problems such as:
$2D$--magnetohydrodynamic turbulence \cite{rr1,rr2,rr3},  $2D$--turbulence
\cite{flo,rahim1}, $c_{p,1}$ models \cite{GK,fl},
gravitationally dressed CFT's \cite{BK}, and some critical disordered
models \cite{CKT,MS}. They play  a role in the so called unifing $W$
algebra \cite{W} and in the description of normalizable zero modes for
string backgrounds \cite{KM,EMN}.
Logarithmic conformal field theories for $D$
dimensional  case ($D>2$) has also been studied \cite{GhK}.

The basic properties of logarithmic operators are that
they form a part of the basis of the Jordan cell
for $L_i$'s, and in the correlator of such fields there
is a logarithmic singularity \cite{Gu,CKT}.
It has been shown that in rational minimal models
two fields with the same dimensions, don't occur \cite{rr2}. In
\cite{RAK} assuming conformal invariance  two-- and
 three--point functions for the case of one or more logarithmic fields
in a block, and one or more  sets of logarithmic fields
have been explicitly calculated. Regarding logarithmic fields
{\it formally} as derivatives of ordinary fields with respect to their
conformal dimension,  $n$--point functions containing
logarithmic fields have been calculated in terms of those of ordinary fields.
These have been done when conformal weights
belong to a discrete set.
In \cite{KAR}, logarithmic conformal field theories with
continous weights have been considered.
The first study of a logarithmic superconformal field theory was carried out
in
\cite{KM,CKLT}.
The WZNW model for SU(2) at the level $k$ is equivalent to the bosonic
sector of the supersymmetric WZNW model at level $k+2$ \cite{VKPR,Fuc}.
In \cite{CKLT} using this equivalence and the results of the WZNW model
for SU(2) at level $k=0$, conformal blocks and OPE's of the
supersymmetric SU$_2$(2) have been obtained. For this supersymmetric
case, some OPE's contain logarithmic terms.

We want to study the general form of correlation
functions of any LSCFT. To do this one should know the general form of
correlation functions of any SCFT. The general form of three--point functions
of SCFT's has been obtained in \cite{Q}. So, at first we construct the
general form of four--point functions of SCFT's.  Super--four--point
functions for the supersymmetric WZNW models are constructed in \cite{Fuc},
in the special case of SU(N) and O(N) symmetry, using superconformal and
super--Kac-Moody Ward identities. It can be easily seen that these
results are in agreement with our general results.

In this article, we construct two--point functions of $N=1$
LSCFT when the Jordan blocks are two or more
dimensional, and when there are one or more Jordan blocks.
For $n$--point functions ($n\geq 3$), the logarithmic correlation functions
can be obtained through formal differentiation of their analogues in ordinary
SCFT's with respect to the superconformal weights.
Using the well known three-point fuctions of SCFT's, one can easily obtain
three--point functions of LSCFT's. We also obtain the general form of $N=1$
SCFT's four--point functions, which can be used for calculating
four--point functions in $N=1$ LSCFT.

\section{Quasi--superprimary operators}
A superprimary operator with conformal weight $\Delta$, is an operator
satisfying
\be\label{1}
[L_n,\Phi (z,\theta )]=[z^{n+1}\partial +(n+1)(\Delta
+{1\over 2}\theta\delta )z^n]\Phi ,
\ee
and
\be\label{2}
[G_r,\Phi (z,\theta )]=\{ z^{r+1/2}\delta -\theta [z^{r+1/2}\partial +
(2r+1)\Delta z^{r-1/2}]\}\Phi .
\ee
Here $L_n$'s and $G_r$'s are the generators of the super Virasoro algebra
satisfying
\be
[L_m,L_n]=(m-n)L_{m+n}+{c\over 12}(m^3-m)\delta_{m+n,0},
\ee
\be\label{4}
[L_m,G_r]=({m\over 2}-r)G_{m+r},
\ee
and
\be
\{ G_r,G_s\} =2L_{r+s}+{c\over 3}(r^2-{1\over 4})\delta_{r+s,0}.
\ee
Throughout this article, the subscripts of $G$'s are half--integers, so that
we are studying the Neveu--Schwarz sector. The superfield $\Phi$ is written as
\be
\Phi (z,\theta )=\phi (z)+\theta\psi (z),
\ee
and
\be
\partial :={\partial\over{\partial z}},\qquad\qquad
\delta :={\partial\over\partial\theta}.
\ee
These, in fact, define a superprimary chiral operator. One can similarly
define a complete superprimary
operator $\Phi (z,\bar z,\theta ,\bar\theta )$
through
\be
\Phi (z,\bar z,\theta ,\bar\theta )=\phi (z,\bar z)+\theta\psi (z)+
\bar\theta\bar\psi (\bar z)+\theta\bar\theta F(z,\bar z),
\ee
relations (\ref{1}) and (\ref{2}), and obvious analogous relations with
$\bar L_n$'s and $\bar G_r$'s.

Now, suppose that the component field $\phi (z)$ has a logarithmic
counterpart (or quasi--primary field) $\phi'(z)$:
\be\label{3}
[L_n,\phi' (z,\theta )]=[z^{n+1}\partial +(n+1)\Delta )z^n]\phi'
+(n+1)z^n\phi (z).
\ee
Our aim is show that $\phi'$ is  the first component of a superfield, which
is the {\it formal} derivative of the superfield $\Phi (z,\theta )$. To do
so, define the fields $\psi'_r$ through
\be
[G_r,\phi'(z)]=:z^{r+1/2}\psi'_r(z).
\ee
Acting on both sides with $L_m$ using the Jacobi identity, and using
(\ref{3}), (\ref{1}), and (\ref{4}), we have
\be
[L_m,\psi'_r(z)]=({m\over 2}-r)z^m(\psi'_{m+r}-\psi'_r)+[z^{m+1}\partial
+(m+1)(\Delta +{1\over 2})z^m]\psi'_r +(m+1)z^m\psi .
\ee
Demanding
\be
[L_{-1},\psi'_r(z)]=\partial\psi'_r(z),
\ee
it is seen that
\be
\psi'_r=\cases{\psi',&$r\geq -1/2$\cr \psi'',&$r\leq -3/2.$\cr}
\ee
Then, equating $[L_1,\psi'_{-5/2}]$ and $[L_1,\psi'_{-3/2}]$, we arrive at
\be
\psi''=\psi'.
\ee
So we have one welldefined field $\psi'$ satisfying
\be
[G_r,\phi'(z)]=z^{r+1/2}\psi'(z),
\ee
and
\be
[L_m,\psi'(z)]=[z^{m+1}\partial+(m+1)(\Delta +{1\over 2})z^m]\psi'
+(m+1)z^m\psi .
\ee
To the end, one can calculate $\{ G_r,\psi'(z)\}$ through the Jacobi
identity. The result is
\bea
\{ G_r,[G_r,\phi']\}&=&{1\over 2}[\{ G_r,G_r\} ,\phi'],\cr
                    &=&[L_{2r},\phi'],\cr
                    &=&[z^{2r+1}\partial +(2r+1)\Delta z^{2r}]\phi' +
                       (2r+1)z^{2r}\phi,
\eea
or
\be
\{ G_r,\psi'\} =[z^{r+1/2}\partial +(2r+1)\Delta z^{r-1/2}]\phi'
+(2r+1)z^{r-1/2}\phi .
\ee
Now, combining $\phi'$ and $\psi'$ in the superfield
\be
\Phi'(z,\theta ):=\phi'(z)+\theta\psi'(z),
\ee
and using the action of super Virasoro generators on component fields
$\phi'$ and $\psi'$ to write the action of super Virasoro generators on the
superfield $\Phi'$, we arrive at
\be\label{5}
[L_n,\Phi'(z,\theta )]=[z^{n+1}\partial +(n+1)(\Delta
+{1\over 2}\theta\delta )z^n]\Phi'+(n+1)z^n\Phi ,
\ee
and
\be\label{6}
[G_r,\Phi'(z,\theta )]=\{ z^{r+1/2}\delta -\theta [z^{r+1/2}\partial +
(2r+1)\Delta z^{r-1/2}]\}\Phi'-(2r+1)z^{r-1/2}\theta\Phi .
\ee
It is easy to see that (\ref{5}) and (\ref{6}) are formal derivatives of
(\ref{1}) and (\ref{2}) with respect to $\Delta$:
\be
\Phi' ={{\d\Phi}\over{\d\Delta}}.
\ee
We call the field $\Phi'$ a quasi--superprimary field, and the two
superfields $\Phi$ and $\Phi'$ a two dimensional Jordanian block of
quasi--primary fields. This has an obvious generalisation to $m$ dimensional
Jordanian blocks:
\be\label{7}
[L_n,\Phi^{(i)}]=[z^{n+1}\partial +(n+1)(\Delta+{1\over 2}\theta\delta )z^n]
\Phi^{(i)}+(n+1)z^n\Phi^{(i-1)},\qquad 1\leq i\leq m-1,
\ee
and
\be\label{8}
[G_r,\Phi^{(i)}]=\{ z^{r+1/2}\delta -\theta [z^{r+1/2}\partial +
(2r+1)\Delta z^{r-1/2}]\}\Phi^{(i)}-(2r+1)z^{r-1/2}\theta\Phi^{(i-1)},\qquad
1\leq i\leq m-1,
\ee
where $\Phi^{(0)}$ is a superprimary field. It is easy to see that (\ref{7})
and (\ref{8}) are satisfied through the formal relation
\be
\Phi^{(i)}={1\over{i!}}{{\d^i\Phi^{(0)}}\over{\d\Delta^i}}.
\ee

\section{Green functions of the Jordanian blocks}
Consider two Jordanian blocks of quasi--superprimary fields $\Phi_1$ and
$\Phi_2$, with the same conformal weight $\Delta$, which are $p$- and
$q$-dimensional, respectively. Invariance of $<\Phi^{(i)}_1(z_1,\theta_1)
\Phi^{(j)}_2(z_2,\theta_2)>$ with respect to $L_{-1}$ and $G_{-1/2}$ shows
that the correlator depends only on
\be
z_{12}:=z_1-z_2-\theta_1\theta_2.
\ee
It is thus sufficient to calculate $<\phi^{(i)}_1(z_1)\phi^{(j)}_2(z_2)>$
to obtain the correlator of the superfields. From \cite{RAK}, however, we
have
\be\label{9}
<\phi^{(i)}_1(z_1)\phi^{(j)}_2(z_2)>=\cases{(z_1-z_2)^{-2\Delta}
     \sum_{k=0}^{i+j-n}{{(-2)^k}\over{k!}}a_{n-k}[\log (z_1-z_2)]^k,
     &$i+j\geq\hbox{max}(p,q)$\cr 0,&$i+j<\hbox{max}(p,q),$\cr}
\ee
as $\phi_1$ and $\phi_2$ are Jordanian blocks of quasi--primary fields. From
(\ref{9}) we have
\be\label{19}
<\Phi^{(i)}_1(z_1,\theta_1)\Phi^{(j)}_2(z_2,\theta_2)>=\cases{
(z_{12})^{-2\Delta}\sum_{k=0}^{i+j-n}{{(-2)^k}\over{k!}}a_{n-k}
[\log (z_{12})]^k,&$i+j\geq\hbox{max}(p,q)$\cr 0,&$i+j<\hbox{max}(p,q).$\cr}
\ee
It is also obvious that such a correlator is nonzero, only if the weights of
the blocks are identical.

The three point functions of Jordanian blocks can be easily obtained by
formal differentiation of the three point functions of the primary
superfield \cite{RAK}. The genral form of the latter has been obtained in
\cite{Q}:
\be
<\Phi_1(z_1,\theta_1)\Phi_2(z_2,\theta_2)\Phi_3(z_3,\theta_3)>=
\prod_{i<j}(z_{ij})^{\Delta -2\Delta_i-2\Delta_j}(a+bW),
\ee
where
\be
\Delta :=\sum_i\Delta_i,
\ee
\be\label{11}
W:={{\theta_1z_{23}-\theta_2z_{13}+\theta_3z_{12}+\theta_1\theta_2\theta_3}
\over{\sqrt{z_{12}z_{13}z_{23}}}},
\ee
and $a$, $b$ are two undetermined constants ($b$ is Grassman valued). To
obtain $<\Phi^{(i)}_1\Phi^{(j)}_2\Phi^{(k)}_3>$, one only needs to perform
differentiation $i$ times with respect to $\Delta_1$, $j$ times with respect
to $\Delta_2$, and $k$ times with respect to $\Delta_3$. In this process,
$a$ and $b$ are also treated formally as functions of $\Delta$'s, so that
each differentiation introduces two more undetermined constants.

The process of obtaining four point functions is exatly the same. First, one
must obtain the general four point functions of super primaryfields. To do
so, one observes that the combination
\be
Y:=\prod_{i<j}(z_{ij})^{\Delta/3-\Delta_i-\Delta_j}
\ee
satisfies the equations obtained through the action of the OSP(2$|$1)
subalgebra of the super Virasoro algebra ($L_{\pm 1}$, $L_0$, $G_{\pm 1/2}$)
on the correlator $<\Phi_1\Phi_2\Phi_3\Phi_4>$. One must now find all
OSP(2$|$1) invariants functions constructed from $z_i$'s and $\theta_i$'s.
The first OSP(2$|$1) invariant is the obvious modification of the anharmonic
ratio $x$, defined as
\be
x:={{(z_1-z_2)(z_3-z_4)}\over{(z_1-z_3)(z_2-z_4)}}.
\ee
It is easy to see that
\be
X:={{z_{12}z_{34}}\over{z_{13}z_{24}}}
\ee
is OSP(2$|$1) invariant. The zeroth order term of $X$ with respect to
$\theta$'s is $x$. Other OSP(2$|$1) invariants are obtained from
combinations odd with respect to Grassman variables. These are analogues of
$W$, eq. (\ref{11}). There are four combinations, but not all of them are
independent. One can show that the cubic terms in $\theta$'s are determined
as linear combinations of the linear terms, and that two linear terms in
$\theta$'s are determined in terms of the other two terms. The coefficients of
these expansions are functions of $X$. The last
OSP(2$|$1) invariant is an even function of the Grassman variables, without
a term of zeroth Grassman order. One can find it to be
\be
V:={{\theta_1\theta_2z_{34}}\over{z_{13}z_{24}}}
  +{{\theta_3\theta_4z_{12}}\over{z_{13}z_{24}}}
  +{{\theta_1\theta_4z_{23}}\over{z_{13}z_{24}}}
  +{{\theta_2\theta_3z_{14}}\over{z_{13}z_{24}}}
  -{{\theta_1\theta_3}\over{z_{13}}}
  -{{\theta_2\theta_4}\over{z_{24}}}
  +{{3\theta_1\theta_2\theta_3\theta_4}\over{z_{13}z_{24}}}.
\ee
So, the most general form of the four point function of superprimary fields
is
\be \label{four}
<\Phi_1\Phi_2\Phi_3\Phi_4>=Y(a+b_1W_{234}+b_4W_{123}+cV),
\ee
where $a$, $b_1$, $b_4$, and $c$ are undetermined functions of $X$ ($b_1$
and $b_4$ are Grassman valued), and
\be
W_{ijk}:={{\theta_iz_{jk}-\theta_jz_{ik}+\theta_kz_{ij}+
\theta_i\theta_j\theta_k}\over{\sqrt{z_{ij}z_{ik}z_{jk}}}}.
\ee
It is easily seen that the results obtained in \cite{Fuc} are in
agreement with the general result (\ref{four}).

Differentiating (\ref{four}) with respect to the
weights, one obtains the most general form of the four point functions of
Jordanian blocks. Once again, one must treat the undetermined functions as
functions of the weights and introduce new functions in each
differentiation.

\newpage

\end{document}